\documentclass[12pt]{article}

\usepackage{amsmath}
\usepackage{amssymb}

\newcommand{\om}{\omega}
\newcommand{\prt}{\partial}
\newcommand{\rot}{\mathrm{rot}}
\newcommand{\di}{\mathrm{div}}

\begin{document}

\title{Topological soliton in magnetohydrodynamics
\footnote{Zh. Eksp. Teor. Fiz. {\bf 82} (1), 117--124 (1982) [Sov. Phys. JETP
{\bf 55,} No.~1, 69--73 (1982)]}}

\author{ A.M. Kamchatnov\\
\small\it Institute of Spectroscopy, Russian Academy of Sciences,
Troitsk, Moscow Region, 142190 Russia
}

\maketitle

\begin{abstract}
We use the Hopf mapping to construct a magnetic configuration consisting of
closed field lines, each of which is linked with all the other ones. We obtain
in this way a solution of the equations of magnetohydrodynamics of an ideal
incompressible fluid with infinite conductivity, which describes a localized
topological soliton.
\end{abstract}

\section{Introduction}

Solutions of physical equations which have non-trivial topological properties
have been studied for already more than five years.   As examples we may give the
``monopole" [1,2] and the ``instanton" [3] in gauge field theories and the
``pseudoparticle" in a two-dimensional isotropic ferromagnet [4].  All these
solutions are characterized by some topological index:  the magnetic charge of
the monopole and the number of pseudoparticles in the ferromagnet are equal to the
degree of mapping of a two-dimensional sphere onto a two-dimensional sphere, the
number of instantons is equal to the Pontryagin index of the mapping of the SU(2)
group onto the three-dimensional sphere.  In each case one can write this index
as a volume integral of some ``topological charge density". In this connection
attention is drawn to the integral of motion [5]
\begin{equation}\label{1}
    I=\int\mathbf{A}\rot\mathbf{A} d^3x
\end{equation}
($\mathbf{A}$ is the vector potential) which has been known for a long time in
the magnetohydrodynamics of a perfectly conducting fluid and which is called the
helicity of the magnetic field.   Its topological nature is already indicated by
the fact that no characteristics of the medium in which the magnetic field is
present enter into (1).   It has also been shown (see Refs. [6,7]) that if two
field line tubes are linked the integral (1) is proportional to their linkage
coefficient, i.e., the number of times which one tube is twisted around the other one.

It is thus clear that the helicity is a topological characteristic of the magnetic
field.   This topological nature of it is completely revealed if we note that (1) is
the same as the Whitehead integral for the Hopf invariant which characterizes
topologically different mappings of the three-dimensional sphere $S^3$ onto the
two-dimensional $S^2$ [8--10]. The topological meaning of the Hopf invariant is simple:
it is equal to the linkage coefficient of the curves in $S^3$ which are the originals
of different points of $S^2$.   Hence follows also a more constructive conclusion:
knowing the mapping $S^3\to S^2$ with a non-zero Hopf invariant, and the simplest
such mapping was constructed by Hopf himself, we can find the vector field $\mathbf{A}$
corresponding to it and, then, the magnetic field $\mathbf{H} = \rot \mathbf{A}$
with non-zero helicity.   The magnetic field lines of this field will be closed and
each of them is linked with any other one.  In the present paper we construct one
such magnetic field configuration and we study its properties in magnetohydrodynamics.

 \section{Stereographic projection and Hopf mapping}

We establish first of all the connection between the physical space $R^3$ and the sphere
$S^3$.   Equation (1) implies that the field $\mathbf{H}$ decreases sufficiently fast at
infinity so that the helicity $I$ is a gauge-invariant quantity: adding to $\mathbf{A}$
the gradient of any function does not change $I$ as the additional term after integration
by parts gives a surface term which does not contribute because $\mathbf{H}$ decreases
rapidly, and a volume term which vanishes because $\di \mathbf{H}= 0$.   If the other
physical conditions at infinity are also unique (say, we consider a homogeneous isotropic
medium) we may assume that the Euclidean three-dimensional space $R^3$ is supplemented
by a point at infinity.   Such a ``compacted" space becomes topologically equivalent to
the three-dimensional sphere $S^3$.   If we embed $S^3$ in the four-dimensional Euclidean
space with coordinates ($u_\mu;  \mu= 1,\, 2,\, 3,\, 4$) so that
$S^3 = \{u_\mu: u_\mu^2= 1\}$ we can establish the connection between $R^3$ and $S^3$
by the stereographic projection
\begin{equation}\label{2}
    x_i=\frac{u_i}{1+u_4},\quad i=1,2,3.
\end{equation}
It is clear that the point at infinity corresponds to the ``south pole" of the sphere
with coordinates ($0, 0, 0, - 1$). The inverse transformation is realized by the formulae
\begin{equation}\label{3}
    u_i=\frac{2x_i}{1+x^2},\quad u_4=\frac{1-x^2}{1+x^2},\quad i=1,2,3,
\end{equation}
where $x^2 = x_i^2$ is the square of the radius vector.

Let there now be in $R^3$ a vector field $\mathbf{A}=(A_1, A_2, A_3)$. We find the
formulae which express the connection between $\mathbf{A}$ and the corresponding vector
field $\widetilde{A}_\mu$ on $S^3$, where we impose on $\widetilde{A}_\mu$
condition that it be tangent to the sphere:
\begin{equation}\label{4}
    u_\mu\widetilde{A}_\mu=0.
\end{equation}
This condition means that both $\mathbf{A}$ and $\widetilde{A}_\mu$ also lie in the
tangent spaces to the appropriate configuration spaces $R^3$ and $S^3$.  We now use
the condition for the invariance of the differential form \footnote{We shall use in this
section the convenient formalism of the theory of exterior forms (see, e.g., Refs. [10, 11]).}
\begin{equation}\label{5}
    \om_A=A_idx_i=\widetilde{A}_\mu du_\mu.
\end{equation}
Since the variables $u_\mu$ are connected through the equation $u_\mu^2 = 1$ for the
sphere in the two expressions on the right-hand side of Eq.~(5), the number of independent
differentials is equal to three.   Taking as the independent variables on $S^3$ the first
three Cartesian coordinates $u_i$ we find from (5)
$$
A_j\frac{\prt x_j}{\prt u_i}=\widetilde{A}_i+\widetilde{A}_4\frac{\prt x_4}{\prt u_i},
\quad i=1,2,3.
$$
Substituting here (2) and using the equation $u_\mu^2=1$, we find three equations
\begin{equation}\label{6}
    \frac{A_i}{1+u_4}+\frac{u_jA_j}{(1+u_4)^2}\frac{u_i}{u_4}=\widetilde{A}_i-
    \frac{u_i}{u_4}\widetilde{A}_4,\quad i=1,2,3,
\end{equation}
which together with (4) are sufficient to express $\widetilde{A}_\mu$ in terms of $A_i$
and vice versa.  As a result we get
\begin{equation}\label{7}
    \widetilde{A}_i=\tfrac12(1+x^2)A_i-x_i(x_jA_j),\quad \widetilde{A}_4=-x_jA_j,
\end{equation}
\begin{equation}\label{8}
    A_i=(1+u_4)\widetilde{A}_i-u_i\widetilde{A}_4,
\end{equation}
where we must also make the coordinate substitutions (2) and (3) respectively
\footnote{ These formulae can also be obtained from the condition that the covariant
derivatives in $R^3$ and $S^3$ are the same (Ref.~[12]).}.

We now describe how, using the known mapping $f: S^3\to S^2$, one must construct the
vector field $\mathbf{A}$ which corresponds to it and which occurs in Eq.~(1) for the
Hopf invariant of this mapping. One must, as was shown by Whitehead (see Ref.~[9] and
also Ref.~[13]), start from the 2-form of the volume on the unit sphere $S^2$.  If
that sphere is embedded in the three-dimensional Euclidean space with coordinates
$\xi_1,  \xi_2, \xi_3$, the 2-form of the volume has the form
\begin{equation}\label{9}
    \om_2=(4\pi)^{-1}(\xi_1d\xi_2\wedge d\xi_3+\xi_2d\xi_3\wedge d\xi_1
    +\xi_3d\xi_1\wedge d\xi_2)
\end{equation}
($\wedge$ is the exterior product sign); the coefficient is here chosen in such a way
that the integral of (9) over the sphere $S^2$ equals to unity.
The mapping $f$ induces a mapping $f^*$ in the opposite direction from the space
of forms on $S^2$ onto the space of forms on $S^3$ so that one can find the 2-form
$f^*\om_2$ on $S^3$.  One can show that any 2-form on $S^3$ can be written in the form
of an exterior differential of some 1-form $\om_1$ where $\om_1$ is determined uniquely
up to a differential of an arbitrary function.   We thus can find a form $\om_1$ such
that $f^*\om_2=d\om_1$.   By using a stereographic projection, we can associate with a
vector field on $S^3$, determined by the form $\om_1$, a vector field $\mathbf{A}$
in $R^3$ which we can use to evaluate the Hopf invariant through Eq.~(1).

We now consider the Hopf mapping $f: S^3 \to S^2$ which has a Hopf invariant equal to
unity (see Refs.~[8,13]):
\begin{equation}\label{10}
    \xi_1=2(u_1u_3+u_2u_4),\quad \xi_2=2(u_2u_3-u_1u_4),\quad
    \xi_3=u_1^2+u_2^2-u_3^2-u_4^2.
\end{equation}
Substituting this formula into (9) we find the form
\begin{equation}\label{11}
    f^*\om_2=\pi^{-1}(du_1 \wedge du_2-du_3 \wedge du_4),
\end{equation}
which is, clearly, the exterior differential of the following form [using the rule
for the evaluation of an exterior differential $d(u_idu_j)=du_l \wedge du_j
=-du_j \wedge du_i$]:
\begin{equation}\label{12}
    \om_1=(2\pi)^{-1}(-u_2du_1+u_1du_2+u_4du_3-u_3du_4).
\end{equation}
The vector field corresponding to this form $\om_1=\om_A$ [see (5)]
\begin{equation}\label{13}
\widetilde{A}=(2\pi)^{-1}(-u_2,u_1,u_4,-u_3)
\end{equation}
satisfies condition (4) so that we can use Eqs.~(8) to find $\mathbf{A}$.
Substituting $\mathbf{A}$, thus found, into (1), indeed, gives $I=1$.
However, it is important for us that $\mathbf{A}$, thus found, can be identified
with the vector potential of a magnetic field with nonzero helicity.

\section{Magnetic field configuration}

We thus find, starting from the vector field (13) and using (8) and (3) a vector
potential in the three-dimensional physical space.  We note that Eqs.~(3) are
clearly written in dimensionless form, i.e., all coordinates $x_i$ refer to some
characteristic dimension $R$.   To change to dimensional units we must make the
substitution $x_i\to x_i/R$, but in order to keep the formulae simple we stay in
this section with dimensionless length units. The dimensional coefficient of
proportionality which fixes the absolute value of the magnetic field strength has
also so far been dropped.  As a result of substituting (13) and (3) into (8) we get,
apart from a proportionality factor
\begin{equation}\label{14}
    A_1=\frac{x_1x_3-x_2}{2(1+x^2)^2},\quad
    A_2=\frac{x_2x_3+x_1}{2(1+x^2)^2},\quad
    A_3=\frac{2x_3^2+1-x^2}{4(1+x^2)^2}.
\end{equation}
Calculating the magnetic field corresponding to this potential we find
\begin{equation}\label{15}
    H_1=\frac{2(x_1x_3-x_2)}{(1+x^2)^3},\quad
    H_2=\frac{2(x_2x_3+x_1)}{(1+x^2)^3},\quad
    H_3=\frac{2x_3^2+1-x^2}{(1+x^2)^3},
\end{equation}
or
\begin{equation}\label{16}
    \mathbf{H}=\rot\,\mathbf{A}/(1+x^2).
\end{equation}
The square of the magnetic field strength equals to
\begin{equation}\label{17}
    H^2=\frac1{(1+x^2)^4},
\end{equation}
so that the absolute magnitude of the magnetic field of the configuration
which we have found is spherically symmetric.

We now find the field lines of the magnetic field (15). The equations of
the lines of force have the form $d\mathbf{x}/dl = \mathbf{H}/H,$ where $dl$
is a line element, or
\begin{equation}\label{18}
    \frac{dx_1}{dl}=\frac{2(x_1x_3-x_2)}{1+x^2},\quad
    \frac{dx_2}{dl}=\frac{2(x_2x_3+x_1)}{1+x^2},\quad
    \frac{dx_3}{dl}=\frac{2x_3^2+1-x^2}{1+x^2}.
\end{equation}
One can easily solve this set of equations if we map it at first on the
sphere $S^3$.   The vector field $\widetilde{H}_\mu$ corresponding to $\mathbf{H}$
is found by using Eqs.~(7) and (2):
\begin{equation}\label{19}
    \widetilde{H}_1=-\tfrac12u_2(1+u_4),\quad \widetilde{H}_2=\tfrac12u_1(1+u_4),\quad
    \widetilde{H}_3=\tfrac12u_4(1+u_4),\quad \widetilde{H}_2=-\tfrac12u_3(1+u_4),
\end{equation}
so that the equations for the lines of force on the sphere $S^3$ have the form
\begin{equation}\label{20}
    \frac{du_1}{d\phi}=-u_2,\quad \frac{du_2}{d\phi}=u_1,\quad
    \frac{du_3}{d\phi}=u_4,\quad \frac{du_4}{d\phi}=-u_3,
\end{equation}
where $d\phi$ is the corresponding line element on $S^3$.   The solution of the
set (20) is clearly:
\begin{equation}\label{21}
    u_1=a\cos(\phi+\phi_1),\quad u_2=\sin(\phi+\phi_1),\quad
    u_3=\sin(\phi+\phi_0),\quad u_4=\cos(\phi+\phi_0),
\end{equation}
where the integration constants $a$ and $b$ are connected through the relation
$a^2 + b^2= 1$.

Again using (2) to change to the physical space we find that the solution of
the set (18) has the form
\begin{equation}\label{22}
    x_1=\frac{a\cos(\phi+\phi_1)}{1+b\cos(\phi+\phi_0)},\quad
    x_2=\frac{a\sin(\phi+\phi_1)}{1+b\cos(\phi+\phi_0)},\quad
    x_3=\frac{b\sin(\phi+\phi_1)}{1+b\cos(\phi+\phi_0)},
\end{equation}
where the $l$-dependence of $\phi$ is found from the differential equation
\begin{equation}\label{23}
    \frac{d\phi}{dl}=\frac2{1+x^2}
\end{equation}
which expresses the well-known connection between the line elements in the
two metrics:   the Euclidean and the
stereographic (see Refs.~[10, 11]). It is clear already from Eqs.~(22) that the
lines of force are closed:   when we change $\phi$ from $0$ to $2\pi$  we
completely traverse it and return to the initial point.   Substituting (22)
into (23) we get
\begin{equation}\label{24}
    \frac{dl}{d\phi}=\frac1{1+b\cos(\phi+\phi_0)},
\end{equation}
so that the length of a line of force is equal to
\begin{equation}\label{25}
    L=\int_0^{2\pi}\frac{d\phi}{1+b\cos\phi}=\frac{2\pi}{(1-b^2)^{1/2}}=
    \frac{2\pi}{|a|}.
\end{equation}

The maximum and minimum values of the radius vector of the points belonging to a
line of force are found from the formulae
\begin{equation}\label{26}
    x_{max}=\left(\frac{1+|b|}{1-|b|}\right)^{1/2},\quad
    x_{min}=1/x_{max}=\left(\frac{1-|b|}{1+|b|}\right)^{1/2}.
\end{equation}
The solution of Eq.~(24) corresponding to the condition $l(0)=0$ has the form
\begin{equation}\label{27}
    l(\phi)=\frac2a\left[\arctan\left(\frac{a}{1+b}\tan\frac{\phi+\phi_1}2\right)-
    \arctan\left(\frac{a}{1+b}\tan\frac{\phi_1}2\right)\right],
\end{equation}
whence we find
\begin{equation}\label{28}
    \tan\frac{\phi+\phi_1}2=\frac{\frac{1+b}a\tan\frac{al}2+\tan\frac{\phi_1}2}
    {1-\frac{1-b}a\tan\frac{al}2\tan\frac{\phi_1}2}.
\end{equation}

Expressing the trigonometric functions in (22) in terms of $\tan[(\phi+ \phi_1)/2]$
and substituting (28) we find the way the equations of the line of force depend on $l$.
We shall not write down the general formulae in view of their complexity, but restrict
ourselves to the case $\phi_1=\phi_0$ as the curves differing only in the difference
$\phi_0-\phi_1$ can be superposed onto one another by a rotation over that angle around
the $x_3$-axis:
\begin{equation}\label{29}
\begin{split}
    x_1&=\frac{(\cos\phi_0+b)\cos al-a\sin\phi_0\sin al}{a(1+b\cos\phi_0)}-\frac{b}a,\\
    x_2&=\frac{(\cos\phi_0+b)\sin al+a\sin\phi_0\cos al}{a(1+b\cos\phi_0)},\quad
    x_3=\frac{b}ax_1.
    \end{split}
\end{equation}
Hence it is clear that the lines of force are plane curves. Evaluation of their curvature gives
\begin{equation}\label{30}
    k=|d^2x/dl^2|=|a|,
\end{equation}
so that the lines of force turn out to be circles of radius $1/|a|$.   This agrees
with Eq. (25) for their length.

Although it follows from the way we have constructed the circles that they are linked,
it is of interest to verify this also directly.   We therefore consider two circles:
$C_1$ and $C_2$ corresponding to values of the parameters $a = b=\sqrt{2}$, $\phi_1=0$
and different values $\phi_0=0$ and $\phi_0=\pi/2$ (one circle is rotated with respect
to the other over $\pi/2$ around the $x_3$-axis).   Their parametric equations have the form
\begin{equation}\label{31}
    \begin{split}
    C_1&=\left\{x_1=\sqrt{2}\cos(l/\sqrt{2})-1,\,x_2=\sin(l/\sqrt{2}),\,x_3=\sin(l/\sqrt{2})\right\},\\
    C_2&=\left\{x_1=-\sin(l/\sqrt{2}),\,x_2=\sqrt{2}\cos(l/\sqrt{2})-1,\,x_3=\sin(l/\sqrt{2})\right\}.
    \end{split}
\end{equation}
The circle $C_1$ lies in the plane $x_2 = x_3$, and the circle $C_2$ in the plane $x_1=-x_3$.
These planes intersect along the line $x_1=-x_2 = -x_3$.   It is clear that if the circles
$C_1$ and $C_2$ are linked, their points of intersection with this line must alternate with
one another.  One easily finds that $C_1$ intersects this line at the points $A_1=(-1,1,1)$
and $B_1 = (1/3, -1/3, -1/3)$,  and $C_2$ in the points $A_2=(1/3, -1/3, -1/3)$ and $B_2=(-1, 1, 1)$.
The point $A_2$ lies between the points $A_1$ and $B_1$ and the point $B_2$ outside the
section $(A_1, B_1)$ so that these pairs of points alternate on the line $x_1= — x_2 = -x_3$
and the circles $C_1$ and $C_2$ are linked.

When the parameter $\phi_0$ changes from $0$ to $\pi/2$ the circle is shifted in space
from $C_1$ to $C_2$ covering a surface with boundaries $C_1$ and $C_2$ which can be obtained
by joining two ends of a strip after twisting it over $360^o$.  It is known (and one can
easily verify this experimentally) that if one cuts such a strip along its boundaries
following a closed line it falls apart into two such strips which are linked.   Continuing
this cutting exercise we shall obtain ever narrower strips which are linked with one another.
It thus becomes clear that all circles forming the original strip are linked with one another.

When the parameter $\phi_0$ changes from $0$ to $2\pi$ the circle describes a closed surface
(a torus obtained from a cylinder which is twisted $360^o$ before it ends are joined)
which is bounding a ``plait" of
closed lines of force.   The lines of force thus lie on toroidal surfaces which are
imbedded one into another, and are circles, each of which is linked with all the others.

We now consider a physical system in which the magnetic field configuration which we have
described can be realized.

\section{Magnetohydrodynamic soliton}

We change to dimensional units so that the magnetic field (15) takes the form
\begin{equation}\label{32}
    \mathbf{H}=\frac{H_0R^4}{(R^2+x^2)^3}\{2R[\mathbf{k}\times\mathbf{x}]+
    2(\mathbf{k}\cdot\mathbf{x})\mathbf{x}+(R^2-x^2)\mathbf{k}\},
\end{equation}
where $\mathbf{k}$ is the unit vector along the $x_3$-axis, $R$ the size of the soliton,
and $H_0$ the magnetic field strength at the origin.   The square of the magnetic field
strength is equal to
\begin{equation}\label{33}
    H^2=\frac{H_0^2 R^8}{(R^2+x^2)^4}.
\end{equation}
Using Eq.~(1) to evaluate the helicity of the magnetic field we get
\begin{equation}\label{34}
    I=\frac{\pi^2}{16}H_0^2R^4.
\end{equation}
We note that through the mapping $\mathbf{x}\to -\mathbf{x}$ we get an ``antisoliton",
the magnetic field of which differs from (32) in the sign in front of the first term in
the braces, while the helicity (34) also changes sign.

We shall consider a perfectly conducting liquid for which $I$ is an integral of motion.
We also restrict ourselves to the case of an incompressible ideal fluid.   The
equations of magnetohydrodynamics for stationary flow have the form (see, e.g., Ref.~[14])
\begin{equation}\label{35}
    \begin{split}
    &\di\,\mathbf{H}=0,\quad \di\,\mathbf{v}=0,\quad \rot\,[\mathbf{v}\times\mathbf{H}]=0,\\
    &(\mathbf{v}\nabla)\mathbf{v}=-\frac1{\rho}\nabla\left(p+\frac{H^2}{8\pi}\right)+
    \frac1{4\pi\rho}(\mathbf{H}\nabla)\mathbf{H}.
    \end{split}
\end{equation}
They are clearly satisfied (see Ref.~[15]) when the fluid moves along the magnetic
field lines of force with a velocity
\begin{equation}\label{36}
    \mathbf{v}=\pm\frac{\mathbf{H}}{(4\pi\rho)^{1/2}}
\end{equation}
while the pressure satisfies the equation
\begin{equation}\label{37}
    p+\frac{H^2}{8\pi}=p_\infty=\mathrm{const}.
\end{equation}
Thus, Eqs.~(32), (36), and (37) give an exact solution of the equations of
magnetohydrodynamics which describes a localized topological soliton.

We evaluate the soliton energy
\begin{equation}\label{38}
    E=\int\left(\frac{\rho v^2}2+\frac{H^2}{8\pi}\right)d^3x=\frac{\pi}{32}H_0^2R^3.
\end{equation}
For a physical interpretation of topological solitons we must bear in mind that they
are metastable states, the energy of which is higher than the energy of a state at
complete equilibrium.   It is thus necessary for the stability of a soliton, at any rate,
that there does not exist such a continuous deformation at which its energy diminishes
while the topological invariant is conserved.   Comparison of (38) and (34) shows that
$$
E\propto\frac{I}R,
$$
so that the soliton can diminish its energy for constant $I$ by increasing its radius.
However, in the case considered there is yet another integral of motion---the angular
momentum [we take the $+$ sign in Eq. (36)]
$$
\mathbf{M}=\rho\int[\mathbf{x}\times\mathbf{v}]d^3x=\frac12(\rho\pi^3)^{1/2}H_0R^4
\mathbf{k},
$$
which stabilizes the ``spreading" of the soliton (cf. the remarks about ``collapse"
of solitons in Refs.~[16, 17]).

The radius $R$ and the field $H_0$ are completely determined by the two conserved
quantities $I$ and ${M}$:
\begin{equation}\label{39}
    R=(M^2/4\pi\rho I)^{1/4},\quad H_0=8(\rho/\pi)^{1/2}I/M.
\end{equation}
For given $I$, the $M$-dependence of the energy has a specific decreasing spectrum,
$E\propto M^{-1/2}$.  One must, however, bear in mind that $I$ and $M$ are not
completely independent quantities.  As the pressure is always positive, it follows
from (37) and (33) that
\begin{equation}\label{40}
    H_0^2\leq8\pi p_\infty
\end{equation}
and thus, according to (39), $I$ and $M$ must satisfy the thermodynamic inequality
\begin{equation}\label{41}
    \frac{I}M\leq\frac{\pi}2\left(\frac{p_\infty}{2\rho}\right)^{1/2}.
\end{equation}

For a given external pressure $p_\infty$ the radius and energy of the soliton
satisfy thus the inequalities
\begin{equation}\label{42}
R\geq(2\pi^4\rho p_\infty)^{-1/8}M^{1/4},\quad
E\leq(\pi^4p_\infty^5/2^{19}\rho^3)^{1/8}M^{3/4}.
\end{equation}
Combining these inequalities [or substituting (40) into (38)] gives
\begin{equation}\label{43}
    E\leq(\pi/2)^2p_\infty R^3
\end{equation}
which is essentially the same as the well known inequality $E<3pV$ which follows
from the fact that the trace of the energy-momentum tensor is positive (see Refs.~[18,19]).

The magnetic field of the soliton (32) is produced by currents which circulate along
closed lines with a density
$$
\mathbf{j}=\frac{c}{4\pi}\rot\mathbf{H}=\frac{c}{2\pi}\frac1{R^2+x^2}
(2R\mathbf{H}+\mathbf{H}\times\mathbf{x}).
$$
These currents are conserved since we neglect dissipative processes.  When account
is taken of the finite conductivity $\sigma$, magnetic field diffusion occurs.
The considerations given here are applicable if the hydrodynamic velocities dominate
the diffusion velocities, i.e., when $(\nu_mR/v)^{1/2}\ll R$, $\nu_m=c^2/4\pi\sigma$
is the magnetic viscosity, or
\begin{equation}\label{44}
    \mathrm{Re}_m=\frac{vR}{\mu_m}\sim\frac{v^2}{c^2}\frac{\sigma R}v\gg1,
\end{equation}
the magnetic Reynolds number must be much larger than unity.  When this criterion is
satisfied, the condition $\sigma R/v\gg1$ that the displacement current is negligible
(see Ref.~[14]), which is assumed to be true in magnetohydrodynamics, is satisfied
automatically (the displacement current vanishes identically in a stationary case
when there is no dissipation).  We can estimate the lifetime of the soliton by
dividing its energy $E$ by
$$
\frac{dE}{dt}=\frac1{\sigma}\int j^2d^3x\sim\frac{c^2H_0^2R}{\sigma}.
$$
As a result we get
\begin{equation}\label{45}
    t\sim\frac{\sigma R}{c^2}.
\end{equation}
When applying inequality (44) to this problem this means that the lifetime (45) is
much longer than the characteristic time $\sim R/v$ for the motion of a fluid particle
along a line of force.

\section{Conclusion}

The equations of magnetohydrodynamics thus admit of an exact solution which describes
a localized topological soliton.   This kind of solution has already been met with
in the physics of the condensed state (see, e.g., Refs.~[16,17]).  We note here some
difference between the magnetohydrodynamic soliton and, say, a soliton in a
ferromagnet [17].  In a ferromagnet the mapping $S^3\to S^2$ is realized by the order
parameter---the magnetization vector $\mathbf{m}(x)$.   Here the sphere $S^2$ has a
direct physical meaning, namely, it is the configuration space of the vector $\mathbf{m}$.
At the same time the map of a point from $S^2$ has no special physical meaning---it is
the line on which $\mathbf{m}(x)$ takes a constant value and the Hopf invariant
characterizes the linking of such lines.   In magnetohydrodynamics there is no
ordering parameter and the sphere $S^2$ has a completely arbitrary character:   its
points merely ``number" the magnetic lines of force and the correspondence between the
lines of force and the points on $S^2$ is established by the Hopf mapping $S^3\to S^2$.
This mapping is not realized in such an apparent manner as in the case of a soliton
in a ferromagnet, but now the maps of the points of $S^2$ have a direct physical
meaning---they are the magnetic lines of force, and the Hopf invariant characterizes
their linking.

I express my gratitude to A.A.~Vedenov, V.G~Nosov, A.L.~Chernyakov, V.R.~Chechetkin, and
V.V.~Yan'kov for discussions of the results of this paper.

\end{document}